# Theory of the Tune Shift Due to Linear Coupling

G. Parzen

## RHIC PROJECT

# Abstract

The presence of skew quadrupole fields will linearly couple the x and y motions. The x and y motions can then be written as the sum of two normal modes . This paper presents analytical perturbation theory results for the tunes of the normal modes. The results for the normal mode tunes are first found correct to lowest order in the skew quadrupole fields. The results are then carried one step further to include the next higher order terms in the skew quadrupole fields. These analytical results show that for the higher order shift in the tune , the important harmonics of the skew quadrupole field are the harmonics near the sum of the tunes. However the harmonics closest to the sum of the tunes do not contribute to the higher order tune splitting, the seperation of the tunes, as they shift the two tunes about equally.This results in a lack of a dominant harmonic for the higher order part of the tune splitting, which complicates the understanding and correction of the higher order part of the tune splitting.

# THEORY OF THE TUNE SHIFT DUE TO LINEAR COUPLING

G. Parzen

Brookhaven National Laboratory
July 1991

## 1. Introduction

The presence of skew quadrupole fields will linearly couple the $x$ and $y$ motions. The $x$ and $y$ motion can then be written as the sum of two normal modes[1] which have the tunes $\nu_1$ and $\nu_2$ which are different from the tune, $\nu_x, \nu_y$, in the absence of the skew quadrupole fields. New beta functions, $\beta_1$ and $\beta_2$, can be defined[2] which are the beta functions of the normal modes and which are different from $\beta_x$ and $\beta_y$, the beta functions of the unperturbed accelerator.

This paper presents analytical perturbation theory results for $\nu_1, \nu_2$. The results for $\nu_1, \nu_2$ are first found correct to lowest order in the skew quadrupole fields. The results for $\nu_1, \nu_2$ are then carried one step further to include the next higher order terms in the skew quadrupole fields. Results for $\beta_1, \beta_2$ will be given in a future paper.

These analytical results show that for the higher order shift in tune the important harmonics of the skew quadrupole field are the harmonics near $\nu_x + \nu_y$. However the harmonics closest to $\nu_x + \nu_y$ do not contribute to the higher order tune splitting, $|\nu_1 - \nu_2|$, as they shift $\nu_1$ and $\nu_2$ about equally. This results in a lack of a dominant harmonic for the higher order contribution of $|\nu_1 - \nu_2|$, which complicates the understanding and correction[3] of the higher order contribution to $|\nu_1 - \nu_1|$.

Analytical results are found for the residual tune splitting which is the $|\nu_1 - \nu_2|$ that remains after the driving term of the nearby difference resonance has been corrected.

## 2. Lowest Order Solution for the Motion and the Tune

The equations of motion can be written as

$$\left(\frac{d^2}{d\theta_x^2} + \nu_x^2\right)\eta_x = b_x(s)\,\eta_y$$

$$\left(\frac{d^2}{d\theta_y^2} + \nu_y^2\right)\eta_y = b_y(s)\,\eta_x$$

$$x = \beta_x^{\frac{1}{2}}\eta_x,\ y = \beta_y^{\frac{1}{2}}\eta_y$$

$$\theta_x = \int ds\,(1/\nu_x\beta_x) = \psi_x/\nu_x \quad (2.1)$$

$$\theta_y = \int ds\,(1/\nu_y\beta_y) = \psi_y/\nu_y$$

$$b_x(s) = \nu_x^2\beta_x\,(\beta_x\beta_y)^{1/2}\,(a_1/\rho)$$

$$b_y(s) = \nu_y^2\beta_y\,(\beta_x\beta_y)^{1/2}\,(a_1/\rho)\ .$$

The skew quadrupole field is described by $a_1(s)$. On the median plane, the field $B_x$ is given by

$$B_x = -B_0\,a_1\,x\ ,$$

where $B_0$ is the main dipole field. $\rho$ is the radius of curvature in the main dipole.

To simplify the solutions of Eq. (2.1), we introduce $\zeta_x$ and $\zeta_y$ such that

$$\eta_x = \zeta_x + \text{c.c.}$$
$$\eta_y = \zeta_y + \text{c.c.} \quad (2.2)$$

$\zeta_x$ and $\zeta_y$ also satisfy Eq. (2.1). In addition, when $a_1 = 0$, the solution for $\zeta_x, \zeta_y$ is

$$\zeta_x = A\exp(i\nu_x\theta_x)$$
$$\zeta_y = B\exp(i\nu_y\theta_y) \quad (2.3)$$

We are looking for a solution of Eq. (2.1) which is valid when $\nu_x, \nu_y$ are close to the coupling resonance $\nu_x - \nu_y = p$, $p$ being some integer. The solution for $\zeta_x, \zeta_y$ will be assumed to have the form

$$\zeta_x = A_s\exp(i\nu_{x,s}\theta_x) + \sum_{r\neq s}A_r\exp(i\nu_{x,r}\theta_x),$$

$$\zeta_y = B_s\exp(i\nu_{y,s}\theta_x) + \sum_{r\neq s}B_r\exp(i\nu_{y,r}\theta_y)\ . \quad (2.4)$$

$$\nu_{x,s} - \nu_{y,s} = p\ .$$

The $A_r$ are assumed to be small compared to $A_s$, and the $B_r$ small compared to $B_s$. $\nu_{x,s}$, $\nu_{y,s}$ will give the $\nu$-values of the normal modes. The normal mode $\nu$-values are $\nu_1$, $\nu_2$ and we assume $\nu_1 \to \nu_x$ and $\nu_2 \to \nu_y$ when $a_1 \to 0$, then $\nu_{x,s} \to \nu_x$ for the $\nu_1$ mode, and $\nu_{y,s} \to \nu_y$ for the $\nu_2$ mode, when $a_1 \to 0$. The justification for choosing this form for the solutions, and the choice of the $\nu_{x,r}$ and the $\nu_{y,r}$ present will come out of the solution one finds using this form.

The $\nu_{x,r}$ and $\nu_{y,r}$ for $r \neq s$ will be seen to have the form

$$\nu_{x,r} = \nu_{x,s} + n$$
$$\nu_{y,r} = \nu_{y,s} + m \tag{2.5}$$

where $n$, $m$ are integers. This could be assumed from the beginning. An alternative procedure is not to restrict $\nu_{x,r}$ and $\nu_{y,r}$, and to make the $\exp(i\nu_{x,r}\theta_x)$ an orthogonal set by choosing $\nu_{x,r} = (2\pi/T)q$, $q$ is some integer and $T$ is some very large angle, and treating $\nu_{y,r}$ similarly. Putting Eq. (2.4) into Eq. (2.1) and using the orthogonal property, one finds

$$\begin{aligned}
\left(\nu_{x,r}^2 - \nu_x^2\right) A_r &= -2\nu_x \sum_{r'} b_x\left(\nu_{x,r}, \nu_{y,r}\right) B_r \\
\left(\nu_{y,r}^2 - \nu_y^2\right) B_r &= -2\nu_y \sum_{r'} b_y\left(\nu_{y,r}, \nu_{x,r'}\right) A_{r'} \\
b_x\left(\nu_{x,r}, \nu_{y,r'}\right) &= \frac{1}{2T} \int_0^T d\theta_x \beta_x \left(\beta_x \beta_y\right)^{\frac{1}{2}} (a_1/\rho) \exp\left[i\left(-\nu_{x,r}\theta_x + \nu_{y,r'}\theta_y\right)\right], \\
b_y\left(\nu_{y,r}, \nu_{x,r'}\right) &= \frac{1}{2T} \int_0^T d\theta_y \beta_y \left(\beta_x \beta_y\right)^{\frac{1}{2}} (a_1/\rho) \exp\left[i\left(-\nu_{y,r}\theta_y + \nu_{x,r'}\theta_x\right)\right].
\end{aligned} \tag{2.6}$$

In Eq. (2.6) we assume $B_r \ll B_s$, $A_r \ll A_s$ for $r \neq s$ and find the first order results

$$\begin{aligned}
\left(\nu_{x,s}^2 - \nu_x^2\right) A_s &= -2\nu_x b_x\left(\nu_{x,s}, \nu_{y,s}\right) B_s \\
\left(\nu_{y,s}^2 - \nu_y^2\right) B_s &= -2\nu_y b_y\left(\nu_{y,s}, \nu_{x,s}\right) A_s \\
\left(\nu_{x,r}^2 - \nu_x^2\right) A_r &= -2\nu_x b_x\left(\nu_{x,r}, \nu_{y,s}\right) B_s \\
\left(\nu_{y,r}^2 - \nu_y^2\right) B_r &= -2\nu_y b_y\left(\nu_{y,r}, \nu_{x,s}\right) A_s
\end{aligned} \tag{2.7}$$

The first two equations in Eq. (2.7) are homogeneous equations for $A_s$ and $B_s$, and the $\nu$-values $\nu_{x,s}, \nu_{y,s}$ are determined by requiring the matrix of the coefficients of $A_s, B_s$

to vanish. This gives

$$\left(\nu_{x,s}^2 - \nu_x^2\right)\left(\nu_{y,s}^2 - \nu_y^2\right) = 4\nu_x\nu_y|\Delta\nu\left(\nu_{x,s}, \nu_{y,s}\right)|^2$$

$$\Delta\nu\left(\nu_{x,s}, \nu_{y,s}\right) = \frac{1}{4\pi}\int_0^{2\pi} ds\,(\beta_x\beta_y)^{\frac{1}{2}}(a_1/\rho)\exp\left[i\left(-\nu_{x,s}\theta_x + \nu_{y,s}\theta_y\right)\right] \quad (2.8)$$

$$\nu_{x,s} - \nu_{y,s} = p$$

Eq. (2.8) can be simplified by assuming that $\nu_x, \nu_y$ are close to the resonance line $\nu_{x,s} - \nu_{y,s} = p$ and $\nu_{x,s} \simeq \nu_x$ and $\nu_{y,s} \simeq \nu_y$. Keeping terms of lowest order only, one gets

$$\left(\nu_{x,s} - \nu_x\right)\left(\nu_{y,s} - \nu_y\right) = |\Delta\nu\left(\nu_{x,s}, \nu_{y,s}\right)|^2$$

$$\nu_{x,s} - \nu_{y,s} = p \quad (2.9)$$

Eq. (2.9) has two solutions for $\nu_{x,s}, \nu_{y,s}$. We denote by $\nu_1$ the value of $\nu_{x,s}$ that goes to $\nu_x$ when $a_1 \to 0$, and $\nu_2$ the value of $\nu_{y,s}$ that goes to $\nu_y$ when $a_1 \to 0$. The solutions can be written as

$$\nu_1 = \overline{\nu}_x \pm \left\{\left(\frac{\nu_x - \nu_y - p}{2}\right)^2 + |\Delta\nu\left(\overline{\nu}_x, \overline{\nu}_y\right)|^2\right\}^{\frac{1}{2}},$$

$$\nu_2 = \overline{\nu}_y \mp \left\{\left(\frac{\nu_x - \nu_y - p}{2}\right)^2 + |\Delta\nu\left(\overline{\nu}_x, \overline{\nu}_y\right)|^2\right\}^{\frac{1}{2}} \quad (2.10)$$

$$\overline{\nu}_x = (\nu_x + \nu_y + p)/2,\, \overline{\nu}_y = (\nu_y + \nu_x - p)/2$$

For the $\pm$, the + sign is used when $\nu_x > \nu_y + p$ for $\nu_1$ and the opposite sign for $\nu_2$. In $\Delta\nu\left(\nu_{x,s}, \nu_{y,s}\right)$, $\nu_{x,s}$ has been replaced by $\overline{\nu}_x$, and $\nu_{y,s}$ by $\overline{\nu}_y$, which introduces a higher order error that can be neglected.

From Eq. (2.10) one finds

$$|\nu_1 - \nu_2 - p| = 2\left\{\left(\frac{\nu_x - \nu_y - p}{2}\right)^2 + |\Delta\nu\left(\overline{\nu}_x, \overline{\nu}_y\right)|^2\right\}^{\frac{1}{2}} \quad (2.11)$$

$$\nu_1 + \nu_2 = \nu_x + \nu_y$$

## 3. Higher Order Shifts in $\nu_1$ and $\nu_2$

To find a higher order result for $\nu_1$ and $\nu_2$, one has to find higher order equations for $A_s, B_s$ by putting the lower order solution for $A_r, B_r, r \neq s$, given by Eq. (2.7) into Eq. (2.6).

Eq. (2.7) for $A_r, B_r$ can be somewhat simplified by assuming that $\nu_x, \nu_y$ are close to the resonance line $\nu_{x,s} = \nu_{y,s} + p$ so that one can assume that $\nu_{x,s} \simeq \nu_x$ and $\nu_{y,s} \simeq \nu_y$ and then

$$A_r = \frac{-2\nu_x b_x (\nu_{x,r}, \nu_{y,s})}{(n + \nu_x + \nu_y)(n - p)} B_s, n \neq p$$
$$B_r = \frac{-2\nu_y b_y (\nu_{y,r}, \nu_{x,s})}{(n + \nu_x + \nu_y)(n + p)} A_s, n \neq -p$$
(3.1)

where $\nu_{x,r} = \nu_{y,s} + n$ and $\nu_{y,r} = \nu_{x,s} + n$.

Putting these results for $A_r, B_r$ in Eq. (2.6) one finds the improved equations for $A_s, B_s$

$$\left(\nu_{x,s}^2 - \nu_x^2 - \Delta_x\right) A_s = -2\nu_x b_x (\nu_{x,s}, \nu_{y,s}) B_s,$$
$$\left(\nu_{y,s}^2 - \nu_y^2 - \Delta_y\right) B_s = -2\nu_y b_y (\nu_{y,s}, \nu_{x,s}) A_s .$$
(3.2)

$$\Delta_x = 4\nu_x \nu_y \sum_{n \neq p} \frac{|c_n|^2}{(n - \nu_x - \nu_y)(n - p)},$$

$$\Delta_y = 4\nu_x \nu_y \sum_{n \neq -p} \frac{|b_n|^2}{(n - \nu_x - \nu_y)(n + p)}$$

$$b_n = \frac{1}{4\pi\rho} \int ds\, a_1 (\beta_x \beta_y)^{\frac{1}{2}} \exp\left[i\left((n - \nu_y)\theta_x + \nu_y \theta_y\right)\right]$$

$$c_n = \frac{1}{4\pi\rho} \int ds\, a_1 (\beta_x \beta_y)^{\frac{1}{2}} \exp\left[i\left((n - \nu_x)\theta_y + \nu_x \theta_x\right)\right]$$

Eq. (3.2) gives the equation for $\nu_{x,s}$ and $\nu_{y,s}$

$$\left(\nu_{x,s}^2 - \nu_x^2 - \Delta_x\right)\left(\nu_{y,s}^2 - \nu_y^2 - \Delta_y\right) = 4\nu_x \nu_y |\Delta\nu(\nu_{x,s}, \nu_{y,s})|^2$$
$$\nu_{x,s} = \nu_{y,s} + p$$
(3.3)

Eq. (3.3) was obtained by using the result for $A_r, B_r$ which is first order in $a_1$. By iterating Eq. (2.6) one can find a result for $A_r, B_r$ to second order in $a_1$ which will change Eq. (3.3) by replacing $\Delta\nu$ by

$$\Delta\nu \to \Delta\nu + \Delta\nu^{(3)}$$
(3.4)

where $\Delta \nu^{(3)}$ is third order in $a_1$. By going one step further and iterating Eq. (2.6) to find results for $A_r$, $B_r$ to third order in $a_1$ will change Eq. (3.3) by replacing $\Delta_x$, $\Delta_y$ by

$$\begin{aligned} \Delta_x &\to \Delta_x + \Delta_x^{(4)} \\ \Delta_y &\to \Delta_y + \Delta_y^{(4)} \end{aligned} \tag{3.5}$$

where $\Delta_x^{(4)}$, $\Delta_y^{(4)}$ are fourth order in $a_1$. One can write down all these higher order terms. However, the expression Eq. (3.3) keeping terms up to second order in $a_1$ is probably sufficient here.

One should also note that in Eq. (3.3) $\nu_{x,s}$ and $\nu_{y,s}$ also occur implicitly in $\Delta \nu (\nu_{x,s}, \nu_{y,s})$ which complicates the solution of Eq. (3.3) for $\nu_{x,s}$ and $\nu_{y,s}$. Solutions can be found depending on the size of $\Delta \nu$ and the distance from the resonance line $\nu_x = \nu_y + p$.

One interesting case is when a 2 family $a_1$ correction system is used to make $\Delta \nu = 0$, and when $\nu_x, \nu_y$ are very close to the resonance line $\nu_x - \nu_y = p$, so that $\nu_1 = \nu_x$ and $\nu_2 = \nu_y$ with an error that is second order in $a_1$. Very close to the resonance line, so that in Eq. (2.10) $(\nu_x - \nu_y - p)^2 /4$ can be neglected compared to $|\Delta \nu|^2$, then the above can be achieved by making $\Delta \nu (\overline{\nu}_x, \overline{\nu}_y) = 0$ as shown in Eq. (2.10).

This corresponds roughly to the situation when a 2 family $a_1$ correction is used to cancel the driving term of the nearby difference resonance, $\nu_x - \nu_y = p$. In this situation, one can find the shift in $\nu_{x,s}$ and $\nu_{y,s}$ due to the second order $\Delta_x$, $\Delta_y$. Then in Eq. (3.3) $\Delta \nu (\nu_{x,s}, \nu_{y,s})$ is not zero but differs from zero by terms of order $a_1^3$, and thus $|\Delta \nu|^2$ is of order $a_1^6$. For this result, the previous observation, that higher order terms can only change the $\Delta \nu$ term by $\Delta \nu^{(3)}$, a term of third order, is significant. As $|\Delta \nu|^2$ is of order $a_1^6$, one can treat it as being zero, and Eq. (3.3) becomes

$$\left( \nu_{x,s}^2 - \nu_x^2 - \Delta_x \right) \left( \nu_{y,s}^2 - \nu_y^2 - \Delta_y \right) = 0 , \tag{3.6}$$

which gives the normal modes

$$\begin{aligned} \nu_1 &= \nu_x + \frac{1}{2\nu_x} \Delta_x \\ \nu_2 &= \nu_y + \frac{1}{2\nu_y} \Delta_y \ . \end{aligned} \tag{3.7}$$

Thus for the case when $\Delta \nu = 0$ and close to the resonance line, there is a second order in $a_1$ shift in the $\nu$-values given by $\Delta_x/2\nu_x$ and $\Delta_y/2\nu_y$. Eq. (3.2) for $\Delta_x$ and $\Delta_y$ show that the largest second order $\nu$-shifts will come from harmonics in $a_1$ close to $\nu_x + \nu_y$. The driving terms $b_n$ and $c_n$ for $n$ closest to $\nu_x + \nu_y$ contribute most to the second order $\nu$-shifts.

One may also notice that $b_n$, $c_n$, as given by Eq. (3.2), are just the usual stop-band results for the $\nu_x + \nu_y = n$ resonance but evaluated at particular points on the resonance line. $b_n$ corresponds to the point $n - \nu_y, \nu_y$ and $c_n$ to the point $\nu_x, n - \nu_x$. For the $n$-values corresponding to resonance lines closest to the unperturbed $\nu_x, \nu_y$, these points on the resonance are not far apart and the $b_n$ and $c_n$ are about equal. Thus for the $\nu_x + \nu_y = n$ lines closest to the unperturbed $\nu_x, \nu_y$, $\nu_1$ and $\nu_2$ are shifted about equally and these $b_n$, $c_n$ do not contribute much to the residual $|\nu_1 - \nu_2|$. This lack of a dominant harmonic for the residual $|\nu_1 - \nu_2|$ makes the correction of the residual $|\nu_1 - \nu_2|$ more difficult.

Eq. (3.7) has been checked[4] by comparing these results with numerical computations of $\nu_1, \nu_2$. For the case of $\nu_x = \nu_y$ resonance line, $p = 0$, Eq. (3.3) may be solved for $\nu_{x,s}$, $\nu_{y,s}$ and written as

$$\nu_1 = \frac{1}{2}\left(\tilde{\nu}_x^2 + \tilde{\nu}_y^2\right) \pm \left\{\left(\frac{\tilde{\nu}_x^2 - \tilde{\nu}_y^2}{2}\right)^2 + 4\nu_x\nu_y|\Delta\nu\left(\nu_1,\nu_1\right)|^2\right\}^{\frac{1}{2}}$$

$$\nu_2 = \frac{1}{2}\left(\tilde{\nu}_x^2 + \tilde{\nu}_y^2\right) \mp \left\{\left(\frac{\tilde{\nu}_x^2 - \tilde{\nu}_y^2}{2}\right)^2 + 4\nu_x\nu_y|\Delta\nu\left(\nu_2,\nu_2\right)|^2\right\}^{\frac{1}{2}} \quad (3.8)$$

$$\tilde{\nu}_x^2 = \nu_x^2 + \Delta_x, \quad \tilde{\nu}_y^2 = \nu_y^2 + \Delta_y$$

$\nu_1$ is the mode that goes to $\nu_x$ when $a_1 \to 0$, and $\nu_2$ goes to $\nu_y$. For the $\pm$ sign, the $+$ sign is used when $\nu_x > \nu_y$ for $\nu_1$ and the opposite sign for $\nu_2$. One can derive Eq. (3.7) from Eq. (3.8) when $\Delta\nu\left(\overline{\nu},\overline{\nu}\right) = 0$, $\overline{\nu} = \frac{1}{2}(\nu_x + \nu_y)$, and close to the resonance line $\nu_x = \nu_y$.

## 4. $\nu$–Shifts when $\nu_x$, $\nu_y$ are far from the $\nu_x - \nu_y = p$ Resonance

In the derivation of the previous results, $\nu_x, \nu_y$ were assumed to be close to the $\nu_x - \nu_y = p$ resonance line. When $\nu_x, \nu_y$ are far from the resonance line the results are less interesting as the $\nu$–shifts are of higher order and smaller. However, it is interesting to see how the results for the $\nu$ shifts in these two cases will fit together.

Up to Eq. (2.6), the previous derivation will hold when $\nu_x, \nu_y$ are far from the $\nu_x - \nu_y = p$ resonance line. Let us first consider the $\nu_1$ mode where $\nu_1 \to \nu_x$ when $a_1 \to 0$. In this case, it is assumed that not only the $A_r$ are small compared to $A_s$, but also $B_s$ is small.

To lowest order, Eq. (2.7) become

$$\begin{aligned}
\left(\nu_{x,s}^2 - \nu_x^2\right) A_s &= 0 \\
\left(\nu_{x,r}^2 - \nu_x^2\right) A_r &= 0 \\
\left(\nu_{x,r}^2 - \nu_y^2\right) B_r &= -2\nu_y b_y \left(\nu_{y,r}, \nu_{x,s}\right) A_s \\
\nu_{y,r} &= \nu_{x,s} + n \ .
\end{aligned} \qquad (4.1)$$

Thus to lowest order, $\nu_1 = \nu_x$, and the tune shift is a higher order effect in $a_1$. To find the second order shift in $\nu_1$, the result for $B_r$ in Eq. (4.1) is put into Eq. (2.6) and the $A_s$ equation becomes

$$\begin{aligned}
\left(\nu_{x,s}^2 - \nu_x^2\right) A_s &= \overline{\Delta}_x A_s \\
\overline{\Delta}_x &= 4\nu_x \nu_y \sum_n \frac{|c_n|^2}{(n - \nu_x)^2 - \nu_y^2} \\
c_n &= \frac{1}{4\pi\rho} \int ds\, a_1 \left(\beta_x \beta_y\right)^{\frac{1}{2}} \exp\left[i\left((n - \nu_x)\theta_y + \nu_x \theta_x\right)\right] \ .
\end{aligned} \qquad (4.2a)$$

This gives the shift in $\nu_x$,

$$\nu_1^2 = \nu_x^2 + \overline{\Delta}_x \ . \qquad (4.2b)$$

The $\overline{\Delta}_x$ is similar to the $\Delta_x$ in Eq. (3.2) except that we now do not assume that $\nu_x - \nu_y \cong p$ and the sum over $n$ is over all $n$. This result, Eq. (4.2b), can be obtained from Eq. (3.3) if in Eq. (3.3) we assume that

$$\left(\nu_{y,s}^2 - \nu_y^2 - \Delta_y\right) \simeq \left((\nu_x - p)^2 - \nu_y^2\right) \ ,$$

and not replace $\nu_x - \nu_y$ by $p$ in Eq. (3.2) for $\Delta_x$.

In the same way one finds for the $\nu_2$ mode,

$$\begin{aligned}
\nu_2^2 &= \nu_y^2 + \overline{\Delta}_y \\
\overline{\Delta}_y &= 4\nu_x \nu_y \sum_n \frac{|b_n|^2}{(n-\nu_y)^2 - \nu_x^2} \\
b_n &= \frac{1}{4\pi\rho} \int ds\, a_1 (\beta_x \beta_y)^{\frac{1}{2}} \exp\left[i\left((n-\nu_y)\theta_x + \nu_y \theta_y\right)\right] \ .
\end{aligned} \qquad (4.3)$$